\documentclass[pra,aps,twocolumn,floatfix,superscriptaddress,showpacs]{revtex4}
\usepackage{graphicx}
\usepackage{amssymb,amsmath}
\usepackage{hyperref}

\newcommand{\ket}[1]{|#1\rangle}
\newcommand{\bra}[1]{\langle#1|}

\newcommand{\eq}{\begin{equation}}
\newcommand{\fine}{\end{equation}}

\begin{document}

\title{

\bf \LARGE Maximal entanglement and state transfer  \\
  using Arthurs-Kelly type interaction for qubits
}

\author{Subhayan Sahu}
\email{subhayan@terpmail.umd.edu} \affiliation{Department of Physics, Indian Institute of Science,Bangalore 560012, India} 

\author{S. M. Roy}
\email{smroy@hbcse.tifr.res.in} \affiliation{HBCSE,Tata Institute of Fundamental Research, Mumbai 400088, India}

\begin{abstract}
We study entanglement generation between a system qubit and three apparatus qubits using an exactly solvable Arthurs-Kelly type model.
We demonstrate the possibility of generating an EPR-like maximally entangled system-apparatus state, in which the second qubit of the 
usual EPR state is replaced by a three qubit state. We design a very simple protocol to transfer the unknown state of 
the system onto one of the apparatus qubits which can then be sent elsewhere via a quantum channel. This protocol can be seen as an alternative teleportation scheme.
\end{abstract}


\maketitle

\section{Introduction}
The idea of Quantum entanglement was introduced by Schr\"odinger \cite{Schrodinger1935}. The truly non-classical features of 
entanglement drew wide attention after the work of Einstein, Podolsky and Rosen (EPR) \cite{Einstein1935} and John Bell \cite{Bell1964}.  In the last two decades, quantum entanglement has emerged as an important resource for Quantum Information processing tasks, such as Quantum Teleportation \cite{Bennett1993,Vaidman1994,Bouwmeester1997,Furusawa1998,Brassard1998,Braunstein2005,Pirandola2008}, Quantum Key Distribution \cite{Ekert1991}, Quantum Computing \cite{Shor,Shor1997,Raussendorf2001} and Quantum Metrology \cite{Roy2008,Boixo2007,Napolitano2011} . Here we report a very simple method for teleportation of 
an unknown state of a system qubit using an interaction that maximally entangles the qubit with three apparatus qubits.

Entangling interactions have been used previously in quantum measurement theory. Von Neumann introduced the idea of tracking of a system observable by 
using an apparatus observable \cite{VonNeumann1955}: the system interacts with the apparatus for some time such that the apparatus observable has the same expectation value in 
the final state as the system observable in the initial state. The idea was extended  by Arthurs and Kelly \cite{Arthurs1965,Braunstein1991,Stenholm1992,Roy2015} to the joint tracking  of two canonically 
conjugate observables by two commuting apparatus observables. The tracking cannot be noiseless and hence, only approximate joint measurement of 
non-commuting system observables is possible. This yields a joint measurement uncertainty relation \cite{Arthurs1988,Stulpe1988,Busch2007} for conjugate observables.
It has also been  shown that the Arthurs Kelly (AK) interaction can be utilized for remote quantum tomography of continuous variable
 systems \cite{Roy2014}. Extensions of AK 
type measurements have been made to joint measurement of different components of spin observables \cite{Levine1989,Busch1986,Pal2011}.

 Here we consider the Levine et al \cite{Levine1989} AK type measurement interaction  between a system qubit and three apparatus qubits
 such that  the three (mutually non-commuting) spin components of the system qubit are tracked by mutually commuting spin components of the 
 apparatus qubits. We first derive a joint measurement uncertainty relation in Sec. \ref{sec:uncertainty}, and then, in Sec. \ref{sec:entanglement}, show that the interaction can give rise to maximal 
entanglement generation between an unknown system qubit and the apparatus. In Sec. \ref{sec:tele} we outline a 
new protocol to transfer the unknown state of the system qubit using the maximally entangled state. The transferred quantum state could then be sent via a quantum channel, effectively acting as an alternate scheme of teleportation, which doesn't require EPR-sharing or Bell state measurements, and instead involves the AK type interaction and single particle spin measurements. In  Sec. \ref{sec:conclusion} we conclude by discussing possible avenues towards experimental implementation of this result.

\section{Joint measurement uncertainty relation} \label{sec:uncertainty}
In the usual AK model for simultaneous approximate measurement of conjugate variables $q,p$ of a system particle P, 
an interaction proportional to $q P_1+p P_2$ with mutually commuting apparatus variables $P_1$, $P_2$ is assumed. There is extensive literature 
on its experimental implementation in quantum optics where $q,p$ denote conjugate quadratures of photons \cite{Arthurs1965,Braunstein1991,Stenholm1992,Roy2015,Busch2007}. 
Consider now possible generalizations to spin measurements.
 In the quantum model of Levine et al \cite{Levine1989}  the three non-commuting spin components of a spin half particle P are 
coupled with three meter qubits ($A_1$, $A_2$ and $A_3$) via an AK type interaction, 
 
\begin{eqnarray}
H &= K(\sigma^P_x \sigma^{A_1}_z + \sigma^P_y \sigma^{A_2}_z + \sigma^P_z \sigma^{A_3}_z) \nonumber\\
  &= K \sum _{i=1}^3  \sigma^P_i \sigma^{A_i}_z
\end{eqnarray} 

where $\sigma^Q_i$ is the $i^{th}$ Pauli Operator for the particle $Q$, where $Q=P,A_1,A_2$ or $A_3$, and $\sigma^Q_1=\sigma^Q_x,\sigma^Q_2=\sigma^Q_y,\sigma^Q_3=\sigma^Q_z$. 
(This is reminiscent of a system particle $P$ of magnetic moment $\bf M$ interacting classically with apparatus particles $A_1,A_2 $ and $A_3$ with respective 
magnetic moments ${\bf M }^1$, ${\bf M }^2$ and  ${\bf M }^3$ via magnetic moment interactions with a Hamiltonian proportional to 
$ {\bf M}.({\bf M }^1+{\bf M} ^2 +{\bf M} ^3) $.)

Neglecting other interactions during the short interaction time $T$, 
the unitary evolution of the four qubit initial state $\ket {{\bf 0}}$ to the final state $\ket{{\bf T}}$ is given by, 
\begin{equation}
 \ket{{\bf T}} =\hat{U} \ket{{\bf 0}} ,
 \hat{U} = \exp[-iHT].
\end{equation}
The unitary evolution operator can be simplified to give
\begin{equation}\label{akint}
\hat{U} = \cos{\theta}{\bf 1}-\frac{i}{\sqrt{3}}\sin{\theta}\sum _{i=1}^3  \sigma^P_i  \sigma^{A_i}_z
\end{equation}
where ${\bf 1}$ denotes the identity operator and $\theta=\sqrt{3}KT$.
The time evolved meter operators after time T in the Heisenberg picture can be written as 
\begin{eqnarray}
 \sigma^{A_i}_x (T)=\hat{U^{\dagger}}\sigma^{A_i}_x \hat{U}=\cos ^2 {\theta} \sigma^{A_i}_x - \sin {(2\theta)} \sigma^P_i \sigma^{A_i}_y/\sqrt{3} \nonumber \\
+\frac{1}{3} \sin ^2 {\theta}\big(\sigma^{A_i} _x +2 \sigma^{A_i}_y \sum _{j,k=1} ^3 \epsilon _{ijk}\sigma^P_j \sigma^{A_k}_z \big),
\end{eqnarray}
where $ \epsilon _{ijk} $ is the totally antisymmetric symbol with $ \epsilon _{123}=1 $, and 
$\sigma^P_1 = \sigma^P_x$, $\sigma^P_2 = \sigma^P_y$, $\sigma^P_3 =\sigma^P_z$; 
this yields, for example,
\begin{eqnarray}
\bra{{\bf T}}\sigma^{A_1}_x \ket {{\bf T}}=\bra{{\bf 0}}\cos ^2 {\theta} \sigma^{A_1}_x - \sin {(2\theta)} \sigma^P_1 \sigma^{A_1}_y/\sqrt{3} \nonumber \\
+\frac{1}{3} \sin ^2 {\theta}\big(\sigma^{A_1} _x +2 \sigma^{A_1}_y (\sigma^P_2 \sigma^{A_3}_z -\sigma^P_3 \sigma^{A_2}_z ) \big)\ket{{\bf 0}}.
\end{eqnarray}
If we start with an initial state,
\begin{equation}
 \ket{{\bf 0}}=\ket{\psi}\ket{+}^{A_1}\ket{+}^{A_2}\ket{+}^{A_3} \equiv \ket{\psi,+,+,+}
\end{equation}
 
 where $\ket{\psi}$ is the unknown initial state of particle $P$, $\ket{\pm}$ are eigenstates of the Pauli Matrix $\sigma_y$
with eigenvalues $\pm 1$,
\begin{equation}
 \sigma_y \ket{\pm} = \pm \ket{\pm}, \sigma_z\ket{+}=\ket{-},
\end{equation}
we obtain,
\begin{equation}
\bra{{\bf 0}}\Sigma_i \ket{{\bf 0}}= \bra{\psi}\sigma^P_i \ket{\psi},
\end{equation}
where, 
\begin{equation}
 \Sigma_i \equiv -\frac {\sqrt{3}} {\sin {(2\theta)}} \sigma^{A_i}_x (T).
 \end{equation}
 For the variances, 
\begin{eqnarray}\label{eq:variance}
&(\Delta \sigma ^P _i ) ^2 =\bra{\psi} (\sigma ^P _i)  ^2 \ket{\psi}  -\bra{\psi} \sigma ^P _i  \ket{\psi}^2 \nonumber \\
&(\Delta \Sigma_i ) ^2 =\bra{{\bf 0}} \Sigma_i  ^2 \ket{{\bf 0}} -\bra{{\bf 0}} \Sigma_i  \ket{{\bf 0}}^2 
\end{eqnarray}
we have the uncertainty relations,
\begin{eqnarray}\label{eq:uncertainty}
&(\Delta \Sigma_i ) ^2 -(\Delta \sigma ^P _i ) ^2=\frac{3}{\sin ^2 {(2\theta)}} -1 \ge 2 \nonumber\\
&\sum_{i=1}^3 (\Delta \sigma ^P _i ) ^2 =2, \nonumber \\
&\sum_{i=1}^3 (\Delta \Sigma  _i ) ^2 =\frac{9}{\sin ^2 {(2\theta)}}-1 \ge 8 .
\end{eqnarray}
The uncertainty relations in Eqs. \ref{eq:variance} and \ref{eq:uncertainty} in the case of spin measurements are the analogues of the measurement uncertainty relations  in the 
standard AK case of $q,p$ measurements.
The tracking of $\sigma ^P _i$ by $\Sigma_i $ is not noiseless; the minimum noise is achieved for $\theta =\pi/4$.  

\section{Entanglement generation} \label{sec:entanglement}

The time evolved state after time T is, 
\begin{eqnarray}
\ket{{\bf T}}=\cos{\theta} \ket{\psi,+,+,+}-\frac{i \sin{\theta} }{\sqrt{3}} \Big(\sigma_x^P \ket{\psi,-,+,+}\nonumber\\
+\sigma_y^P\ket{\psi,+,-,+}+\sigma_z^P\ket{\psi,+,+,-} \Big).
\end{eqnarray} 
If the system qubit is denoted by 
\begin{equation}
 \ket{\psi }=
\begin{pmatrix}
	a\\
	b
\end{pmatrix}
=a\ket{0}+b\ket{1},
\end{equation}
the above state can be expressed in the following form:
\begin{multline}
\ket{{\bf T}}=\ket{0}\Big(a \cos{\theta}\ket{+++}-i(\sin{\theta}/\sqrt{3})\\ \times \big(b\ket{-++}-i b\ket{+-+}+a\ket{++-}\big)\Big)\\
+\ket{1}\Big(b \cos{\theta}\ket{+++}-i(\sin{\theta}/\sqrt{3})\\ \times \big(a\ket{-++}+i a\ket{+-+}-b\ket{++-}\big)\Big).
\end{multline}
Note that the apparatus states multiplying the system states $\ket{0}$ and $\ket{1}$ are mutually orthogonal if and only if $\cos^2{\theta}=1/4$; 
in that case the above state is expressed in the Schmidt bi-orthogonal form \cite{Schmidt1907}.

The final reduced density matrices for the system qubit $P$ and the three-qubit apparatus $A={A_1,A_2,A_3}$ are,
\begin{equation}
\rho^P = Tr_{\{A_1,A_2,A_3\}}\ket{{\bf T}}\bra{{{\bf T}}} ; \rho^{A} = Tr_{(P)} \ket{{\bf T}}\bra{{{\bf T}}}.
\end{equation}
This yields, 

\begin{equation*}
\rho^P_{00}=\cos^2\theta |a|^2+\frac{1}{3}\sin^2\theta(1+|b|^2)
\end{equation*}
\begin{equation*}
\rho^P_{01}=(\rho^P_{10})^*=(\cos^2\theta -\frac{1}{3}\sin^2\theta)ab^{*}
\end{equation*}
\begin{equation*}
\rho^P_{11}=\cos^2\theta |b|^2+\frac{1}{3}\sin^2\theta(1+|a|^2)
\end{equation*}
The system-apparatus entanglement $E$ is given by the von Neumann entropy of the reduced density matrix of system $P$ or equivalently of the apparatus $A$,
\cite{Bennett1996}
\begin{eqnarray}
 E= -Tr  \rho^P  ln \rho^P = -Tr  \rho^A  ln \rho^A \nonumber \\
 = - \lambda  \ln  \lambda - (1-\lambda)  \ln  (1-\lambda),
 \end{eqnarray}
 where $\lambda$ and $1-\lambda$ are eigenvalues of $\rho^{P}$ which obey
 \begin{equation}
\lambda(1-\lambda)= det \rho^{P}= \frac{2}{9}\sin^2\theta(1+2\cos^2\theta) .\nonumber
\end{equation} 
The entanglement is maximum when $\lambda =1/2$ , i.e. $cos^2\theta=1/4$, 
\begin{equation}
 E  \leq  \ln 2;  E=  \ln 2 \text{ for } \cos^2\theta=1/4.
\end{equation}

For $\cos \theta=1/2$, $\sin \theta =\pm \sqrt{3}/2$, the corresponding maximally entangled final states $\ket{{\bf T}\pm} $ assume the Schmidt bi-orthogonal forms,
\begin{multline}\label{maxentschmdt}
\ket{{\bf T}\pm} =\frac{\ket{0}}{2}\Big(a \ket{+++} \mp i\big(b\ket{-++}\\-i b\ket{+-+}+a\ket{++-}\big)\Big)\\+\frac{\ket{1}}{2}\Big(b \ket{+++}
\mp i\big(a\ket{-++}\\+i a\ket{+-+}-b\ket{++-}\big)\Big).
\end{multline}
These states, are analogous to the two qubit EPR states, with one of the qubits replaced  by three qubits.

 Since $\theta=\sqrt{3}KT$, it is seen that by varying the product of the strength and duration of interaction, such that $\cos \theta=1/2$, 
 $\sin\theta=\pm \sqrt{3}/2$, maximal entanglement between the system and the apparatus can be achieved.
\section{Quantum state transfer with the maximally entangled state}\label{sec:tele}
 
\begin{figure*}
		\includegraphics[width=\textwidth]{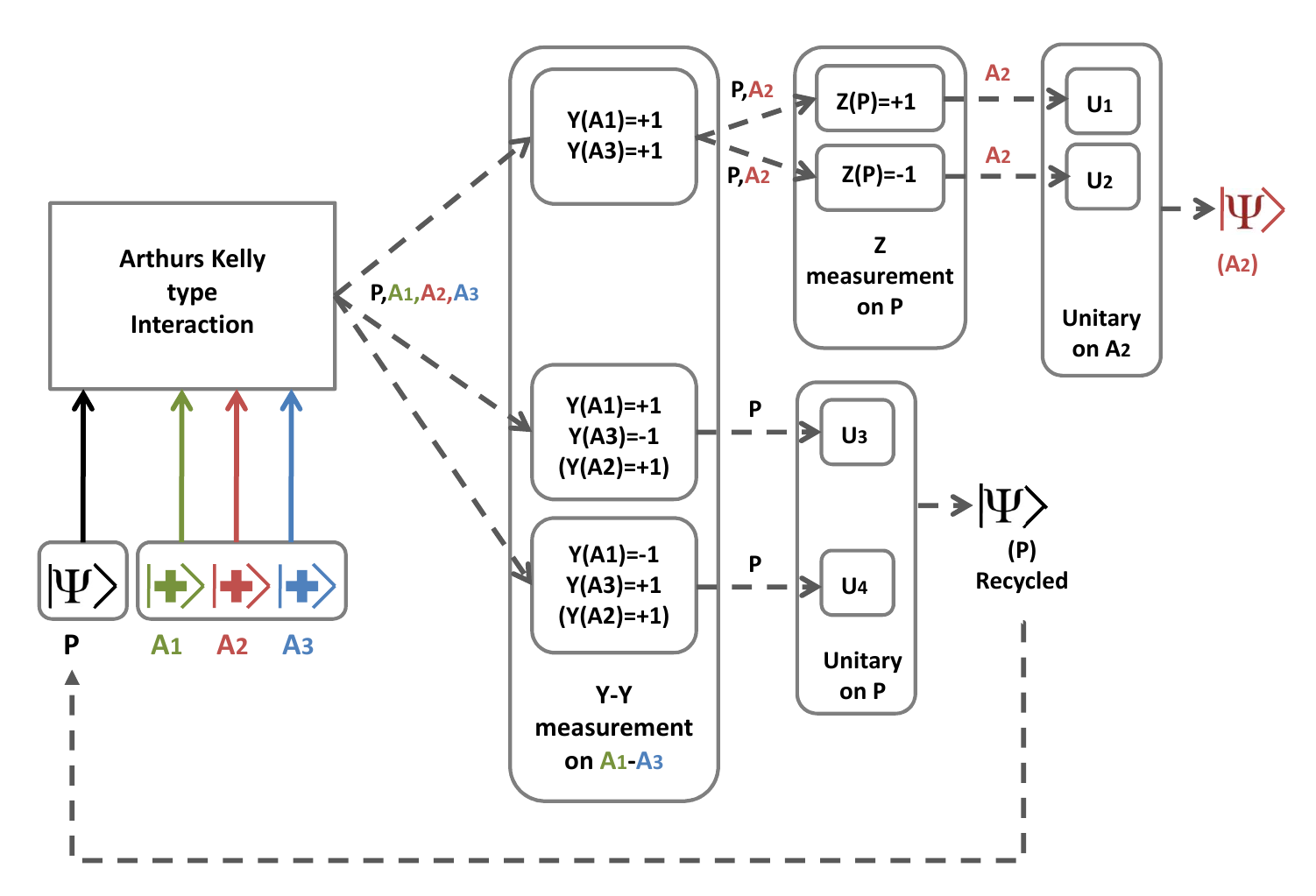}
		\caption{
			A schematic diagram showing the quantum state transfer protocol for the case  $\cos \theta=1/2$, $\sin \theta =+ \sqrt{3}/2$. 
			The qubit P is the system qubit in an unknown state, and the qubits $A_1$,$A_2$ and $A_3$ 
			are suitably prepared apparatus qubits. After an AK type interaction, 
			a  measurement of $\sigma_y ^{A_1}=Y(A_1)$ and $\sigma_y ^{A_3}=Y(A_3)$  separates the emerging particles into 
			three beams with probabilities 1/2, 1/4 and 1/4. The first beam with $Y(A_1)=+1, Y(A_3)=+1$ (probability 1/2) is subjected to 
			a measurement of $\sigma_z ^P=Z(P)$: this yields two beams of $A_2$ particles with $Z(P)=+1$ and $Z(P)=-1$ which are then 
			subjected to the unitary transformations $U_1$ and $U_2$ respectively (see Eq. (\ref {U})) to yield the state  $\ket {\psi}^{A_2}$ 
			which can be teleported through a quantum channel. The second beam with $Y(A_1)=+1, Y(A_3)=-1,Y(A_2)=+1 $ (probability 1/4) is 
			subjected to a unitary transformation $ U_3=\sigma_z $ on the particle $P$,and the third beam $Y(A_1)=-1, Y(A_3)=+1,Y(A_2)=+1 $ 
			(probability 1/4) is subjected to a unitary transformation $ U_4=\sigma_x $ on the particle $P$. In both cases the original 
			state $\ket {\psi}^{P}$ is recovered and recycled to get a fresh sequence of AK intearction and measurements.
                        For $cos \theta=1/2$, $sin \theta = -\sqrt{3}/2$ the only change necessary in the flow chart is that 
                        $U_1\rightarrow \sigma_z U_1$ and $U_2\rightarrow \sigma_z U_2 $.
                        The probability of getting the transferred state $\ket {\psi}^{A_2}$ after 0,1,2,... recycling are $\frac{1}{2}$,$\frac{1}{4}$,$\frac{1}{8}$,...which add up to 1.
  		} \label{fig:fig1}
\end{figure*}

Suppose we wish to transfer the unknown state of the particle $P$ on to one of the apparatus particles, say $A_2$, which can then be teleported.
This might be useful e.g. if  $A_2$ is more easily transported over a quantum channel than $P$ or if  it has a longer lifetime than $P$. If we 
expand the state $\ket{\bf T}$ for general $\theta$ in the basis of the orthogonal states $\ket {\pm}$ for the $A_1,A_3$ particles, we obtain,
\begin{eqnarray}\label{teleport}
\ket{{\bf T}}= \ket{+}^{A_1}\ket{+}^{A_3} \nonumber \\
\Big(\ket{0} (a \cos{\theta}\ket{+}^{A_2}-b(\sin{\theta}/\sqrt{3})\ket{-}^{A_2})\nonumber\\
+ \ket{1} (b \cos{\theta}\ket{+}^{A_2}+a(\sin{\theta}/\sqrt{3})\ket{-}^{A_2}) \Big ) \nonumber\\
+ \ket{+}^{A_1}\ket{-}^{A_3}\ket{+}^{A_2}(-i\sin{\theta}/\sqrt{3}) (a \ket{0} -b \ket{1} ) \nonumber\\
+ \ket{-}^{A_1}\ket{+}^{A_3}\ket{+}^{A_2}(-i\sin{\theta}/\sqrt{3}) (b \ket{0} +a \ket{1} ).
\end{eqnarray}
Strikingly, we see a connection between maximal entanglement and perfect state transfer: 
the coefficients of $ \ket{+}^{A_1}\ket{+}^{A_3}\ket{0} $ and $ \ket{+}^{A_1}\ket{+}^{A_3}\ket{1}$ are states of particle $A_2$ which are unitary 
transforms of the original unknown state of particle $P$, if and only if $ \sin \theta=\pm \sqrt{3}/2$ (maximal entanglement between the system and the apparatus).
For these special values of $\theta$, Eq. \ref{teleport} immediately suggests the following teleportation protocol.
If we make measurements on $\ket{{\bf T}\pm}$ to project it on to the sub-spaces $\ket{+}^{A_1}\ket{+}^{A_3}\ket{0}$ ,and 
$\ket{+}^{A_1}\ket{+}^{A_3}\ket{1}$, 
we obtain respectively the following normalized states of the qubit $A_2$,
\begin{eqnarray}
 2 \bra{0}\bra{+}^{A_1}\bra {+}^{A_3} \ket{{\bf T}\pm}  = a \ket{+}^{A_2}\mp b \ket{-}^{A_2} \nonumber \\
 =-iU_1^{\dagger}(a \ket{0}\pm b\ket{1})^{A_2} \nonumber\\
  2 \bra{1}\bra{+}^{A_1}\bra {+}^{A_3} \ket{{\bf T}\pm} = b\ket{+}^{A_2}\pm a \ket{-}^{A_2}\nonumber \\
 =\pm iU_2^{\dagger}( a \ket{0} \pm b\ket{1})^{A_2},
 \label{state2}
\end{eqnarray}
where $U_1$ and $U_2$ are the unitary transformations
\begin{equation}
\label{U}
U_1=\frac{1}{\sqrt{2}}\begin{pmatrix}
1&-i\\
-1&-i
\end{pmatrix};
U_2=\frac{1}{\sqrt{2}}\begin{pmatrix}
1&i\\
1&-i
\end{pmatrix}
\end{equation}

For general $\theta $,  we can read off the corresponding non-unitary transformations $\hat{U}_1$, $\hat{U}_2$, and see that,
\begin{equation} 
\label{nonunitarity}
 (\hat{U}_1 ^\dagger \hat{U}_1 + \hat{U}_2 ^\dagger \hat{U}_2 )/2 -{\bf 1} = \big (1-(4 /3)\sin ^2 \theta \big ) {\bf 1 },
\end{equation}
where $\bf 1 $ denotes the unit matrix, and the right-hand side gives a quantitative measure of the imperfection of teleportation when  
$ \sin ^2 \theta \neq  3/4$.
 
If $ \sin ^2 \theta =  3/4$, we recover $\ket{\psi}^{A_2}$ by applying the appropriate unitary transforms $U_1$ , $U_2$ if $cos \theta=1/2, \sin\theta=+ \sqrt{3}/2$, 
and the unitary transforms $\sigma_z U_1$ , $\sigma_z U_2$ if $cos \theta=1/2, \sin\theta=- \sqrt{3}/2$.
On the other hand coefficients of  $ \ket{+}^{A_1}\ket{-}^{A_3}\ket{+}^{A_2}$ and $ \ket{-}^{A_1}\ket{+}^{A_3}\ket{+}^{A_2}$ are 
proportional to $U_3 \ket{\psi}^P$ and $U_4 \ket{\psi}^P$, where,
\begin{equation}
 U_3= \sigma_z; U_4= \sigma_x
\end{equation}

 A flowchart of the state transfer protocal is provided in Fig. \ref{fig:fig1} for the case  $\cos \theta=1/2$, $\sin \theta =+ \sqrt{3}/2$.
  A  measurement of $\sigma_y ^{A_1}=Y(A_1)$ and $\sigma_y ^{A_3}=Y(A_3)$ 
  separates the emerging particles into three beams with probabilities 1/2, 1/4 and 1/4 .
  
  The first beam with $Y(A_1)=+1, Y(A_3)=+1$ (probability 1/2) is subjected to a measurement of $\sigma_z ^P=Z(P)$: this 
  yields two beams of $A_2$ particles with $Z(P)=+1$ and $Z(P)=-1$ which are then subjected to the unitary transformations $U_1$ and $U_2$ respectively 
  (see Eq. \ref {U}) to yield the state  $\ket {\psi}^{A_2}$ which is teleported through a quantum channel.
  
  The second beam with $Y(A_1)=+1, Y(A_3)=-1,Y(A_2)=+1$ (probability 1/4) is subjected to a unitary transformation $ U_3=\sigma_z $ on the particle $P$,
  and the third beam $Y(A_1)=-1, Y(A_3)=+1,Y(A_2)=+1 $ (probability 1/4) is subjected to a unitary transformation $ U_4=\sigma_x $ on the particle $P$; in both cases the 
  original state $\ket {\psi}^{P}$ is recovered and recycled to get a fresh sequence of AK intearction and measurements.
  
  For $\cos \theta=1/2$, $\sin \theta = -\sqrt{3}/2$. the only change in the flow chart is that 
\begin{equation}
  U_1\rightarrow U_1'=\sigma_z U_1 ; U_2\rightarrow U_2'=\sigma_z U_2 .
\end{equation}

 The probability of getting the teleported state $\ket {\psi}^{A_2}$ after 0,1,2,... recycling are $\frac{1}{2}$,$\frac{1}{4}$,$\frac{1}{8}$,... which add up to 1.
 
 {\bf Entanglement Swapping.} Suppose now that the particle $P$  sent for AK type interaction, instead of being in a state $\ket {\psi} ^P$ is 
 actually entangled with another particle $R$ in Alice's laboratory, and their joint state is,
  \begin{equation}
   \ket {\phi_1 }^R \ket {\psi_1} ^P + \ket {\phi_2 }^R \ket {\psi_2} ^P.
  \end{equation}
Then, using the linearity of Schrödinger equation, after the state of particle $P$ is teleported to that of $A_2$ in Bob's laboratory, the particle $R$ in 
Alice's lab. becomes entangled with  $A_2$ in Bob's lab. and their joint state is,
 
 \begin{equation}
   \ket {\phi_1 }^R \ket {\psi_1} ^{A_2} + \ket {\phi_2 }^R \ket {\psi_2} ^{A_2}.
  \end{equation}  
  
  The main difference from the usual protocol is that the particle $A_2$ is taken not from a previously prepared EPR pair, but from the final state 
  of the Arhturs-Kelly type interaction.

{\bf Comparison with usual teleportation protocols}. To re-iterate, our proposed scheme of quantum state transfer followed by teleportation, differs from the original teleportation scheme envisaged by Bennett et al \cite{Bennett1993}. The conventional scheme has 4 main steps: (i) An EPR pair E1, E2 is shared by Alice and Bob 
at distant locations. (ii) The system particle P with unknown state is received by Alice and she makes a Bell-state measurement on the joint state of that 
particle and E1 and (iii) communicates the result via a classical channel to Bob; (iv) Bob then makes a unitary transformation on E2 depending on the classical 
information  to replicate the unknown system state.
In the alternative Teleportation scheme reported here, the steps of EPR-pair sharing, Bell projection and classical communication are not necessary; 
instead, the AK entangling interaction and single particle spin measurements are used to ``transfer" an unitary transform of the unknown state 
to an apparatus qubit. The unknown state can be recovered from the apparatus qubit by applying the inverse unitary transform, 
either before or after teleportation of that qubit. Importantly, all the qubits interact in this scheme, while in the original scheme, the qubit to teleported doesn't interact with Bob's qubit. Also, this scheme could be useful, as was mentioned before, if  the apparatus qubit is more easily transported over a quantum channel or if it has a longer lifetime than the original qubit. Another advantage of the present scheme, is that 
single particle spin measurements are much easier than Bell state measurements.

\section{Conclusion and outlook}\label{sec:conclusion}
We have shown that the  Levine et al \cite{Levine1989} AK type interaction can generate maximal entanglement between  a system qubit and three apparatus qubits.
We utilise this to introduce a novel scheme of teleportation which has some advantages over the conventional methods.
The main new task is the realisation of the AK interaction. The technology and experimental realizations of the 
 AK interaction for optical quadratures (such as $q,p$) are widely known in the context of optical 
 Homodyne and Heterodyne  Tomography  (see e.g. \cite{Vogel1989,Braunstein1990,Smithey1993,Leonhardt1995,Yuen1978,Shapiro1979,Yuen1980,Stenholm1992}). The present work provides a concrete motivation to extend the technology to  
 realize the Levine et al \cite{Levine1989} proposal of an AK type interaction between qubits. 
 Possible qubit systems are solid-state nuclear and electronic qubits in diamond \cite{Pfaff2014}, in ion-traps  \cite{Olmschenk}, and polarized photon qubits as in tests of Bell inequalities \cite{Aspect1982}.
Once the teleported state $\ket{\psi}^{A_2}$ is realized, applications to long distance quantum communications via quantum memory and quantum 
repeaters might be possible \cite{Lvovsky2009,Morton2008,Saeedi2013,Nafradi2016}.

Another distinct approach towards practically implementing this interaction for quantum information processing tasks is through quantum simulation \cite{Childs2017,Verstraete2009} of the Levine et al \cite{Levine1989} Hamiltonian. With the advent of Noisy Intermediate-Scale Quantum (NISQ) technology \cite{Preskill2018}, many of which are already available, this could be an interesting direction of future research.   
 

{\bf Acknowledgments.}
S.S. thanks the NIUS program in HBCSE, TIFR and KVPY scholarship program; S. M. R thanks the Indian National Science Academy for an INSA senior scientist grant. The authors thank the referee for pointing out a necessary clarification in nomenclature. 

The authors contributed equally to the research reported in this paper.

\bibliography{SMR}

\end{document}